\documentclass[%
aps,%
prb,%
twocolumn,
superscriptaddress,%
showpacs,%
preprintnumbers,%
byrevtex%
]{revtex4}

\usepackage{ifthen}
\usepackage{color}
\usepackage{amssymb}
\usepackage{amsfonts}
\usepackage{amsmath}
\usepackage[dvips]{graphicx}

\def\ii{\textrm{i}\,\!}

\newcommand{\beann} {\begin{eqnarray*}}
\newcommand{\eeann} {\end{eqnarray*}}
\newcommand{\bea} {\begin{eqnarray}}
\newcommand{\eea} {\end{eqnarray}}
\newcommand{\labs} {\left\vert}
\newcommand{\rabs} {\right\vert}

\newcommand{\lrb} {\left(}
\newcommand{\rrb} {\right)}
\newcommand{\lcb} {\left\{}
\newcommand{\rcb} {\right\}}
\newcommand{\lab} {\left\langle}
\newcommand{\rab} {\right\rangle}

\begin{document}
\date{October 10, 2003}

\title{Giant magnetoresistance of multiwall carbon nanotubes:
\\ modeling the tube/ferromagnetic-electrode burying contact}

\author{S.~Krompiewski}
\affiliation{Institute of Molecular Physics, Polish Academy of Sciences,
PL-60179 Pozna{\'n}, Poland}

\author{R.~Guti{\'e}rrez}
\author{G.~Cuniberti}
\affiliation{Institute for Theoretical Physics, University of Regensburg,
D-93040 Regensburg, Germany}

\begin{abstract}
We report
on the giant magnetoresistance (GMR) of multiwall carbon nanotubes with ultra small diameters.
In particular, we consider the effect of the inter-wall interactions and the lead/nanotube coupling.
 Comparative studies have been performed to show that in the case when all walls
 are well coupled to the electrodes, the so-called inverse GMR can appear.
 The tendency towards a negative GMR depends on the inter-wall interaction and on the nanotube length.  If,
 however, the inner nanotubes are out of contact with one of the electrodes, the GMR
 remains
 positive even for relatively strong inter-wall interactions regardless of the outer nanotube length.
 These results shed additional
light on recently reported experimental data, where an inverse GMR
was found in some multiwall carbon nanotube samples.
\end{abstract}

\maketitle

\section{Introduction}
Carbon nanotubes belong to the most promising new materials for
the future molecular electronics, they are believed to potentially
replace in the near future the silicon-based conventional
electronics.  To illustrate the enormous scientific and
technological progress that has been made since carbon nanotubes
were discovered it is worth to mention new concepts such as: the
room temperature single electron transistor,\cite{PTYGD01}
 the ballistic carbon nanotube field-effect
transistor\cite{JaveyGWLD03} or the non-volatile random access
memory for molecular computing.\cite{RKJTCL00} Recently several
both
experimental\cite{TsukagoshiAA99+JensenNB03,ZhaoMVMS02,KimKPKKKW02}
and
theoretical\cite{MehrezTGWR00,Krompiewski03pss,Krompiewski04jmmm}
papers have been published on spin-dependent electrical transport
in ferromagnetically contacted carbon nanotubes in an attempt to
test their ability to operate as spintronic devices.  It has been
found that carbon nanotubes --though intrinsically nonmagnetic--
reveal quite a considerable GMR effect.  However, it should be
stated in this context that the experimental results are very
diverse, and reflect to a great extent sample-specific features.
The poor reproducibility of experiments on carbon nanotubes, or
more generally on molecular systems, is due to hardly controllable
interface conditions between the molecule and external electrodes.
Incidentally, judging from the conductance value, this difficulty
is
 sometimes successfully overcome
but, to our knowledge, only if nonmagnetic-electrodes are used.
The examples proving this statement are numerous as far as
single wall carbon nanotubes (SWCNTs)
are concerned,\cite{LiangBBHTP01,JaveyGWLD03} and much more seldom in the case of
multiwall carbon nanotubes (MWCNTs) \cite{FrankPWdH98}.
 Despite intensive studies,\cite{KwonT98+SKTL00,CunibertiGG02}
fundamental problems such as (\textit{i}) the internal structure
of nanotubes (especially multiwall ones), (\textit{ii}) the
aforementioned nature of the electrode/nanotube coupling, and
(\textit{iii}) the energy band line-up at the interfaces are still
far from being well understood.  Motivated by these facts, we
study a minimal geometrical model of the MWCNT, which enables us
to define in a unique way the coupling of the MWCNT to the
electrodes.  Having done that, we can restrict ourselves to the
most crucial issues addressed in the present studies, namely: the
inter-wall interactions, and the effect of the breakage of the
contact between the inner wall and one of the transversally
coupled (burying) electrodes.  So far tangential contact
geometries have been mainly studied in connection with STM
experiments.\cite{Tersoff99+ADX00+HanssonS03}
 However in experiments
with ferromagnetic-electrodes, transition metals are usually deposited by evaporation on top
of the carbon nanotube cutting off a segment of the latter, so
that the nanotube finds itself buried in the electrode and the contact geometry can be regarded as transversal.
\begin{figure}[b]
\centerline{\includegraphics[width=.99\linewidth]{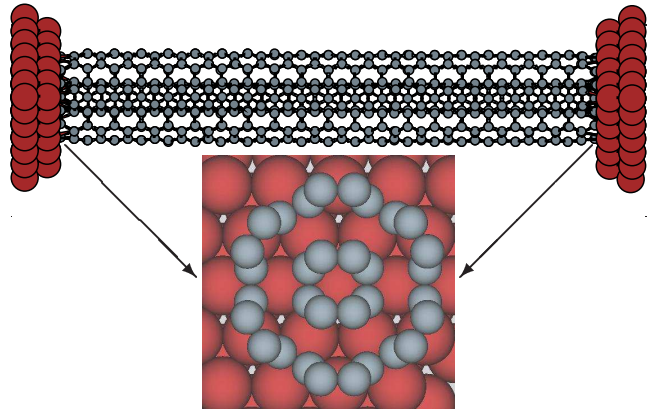}}
\caption{\label{fig:fig1}(Color online) View of the (2,2)@(6,6) carbon nanotube sandwiched between two fcc(111) leads and detail of
the contact region.  What is shown consists of a few ferromagnetic-electrode atoms with the nanotube forming
the so-called extended molecule.  The other parts of the electrodes (not shown) are infinite in all the directions.}
\end{figure}

\begin{figure}[t]
\centerline{\includegraphics[width=.99\linewidth]{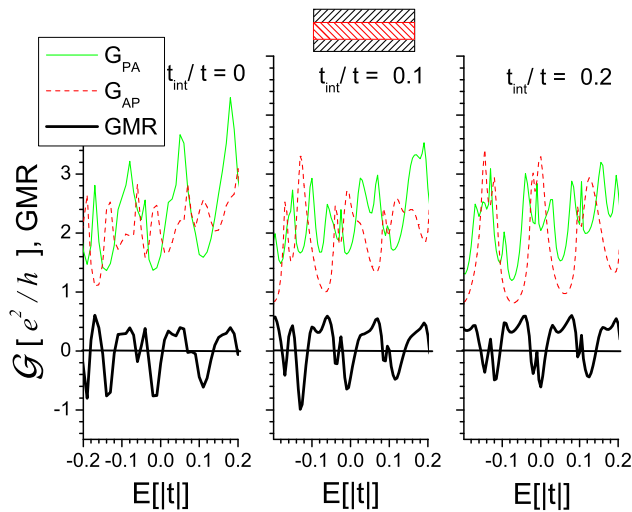}}
\caption{\label{fig:fig2} (Color online) The total conductance for the parallel
(PA) and antiparallel (AP) alignments, as well as the GMR
\textit{vs.}~energy.  The outer nanotube length equals that of the
inner one ($L_{\rm out}$=$L_{\rm in}$=41 carbon rings), the Fermi
energy is equal to $E_{\rm F}=0$ and $t = -2.66$.  Left panel: there are no
inter-wall interactions ($t_{\rm int} /t =0$).  Middle and right
panels: the inter-wall interactions are included, $t_{\rm int} /
t= 0.1$ and $0.2$, respectively. }
\end{figure}
\section{Modeling premise}
Our system consists of a double-wall carbon nanotube (DWCNT)
perfectly contacted to semiinfinite electrodes having an fcc(111)
geometry: every carbon atom at the interface has exactly three
nearest neighbors in the electrode (see Fig.~\ref{fig:fig1}). Such
a construction is possible in the case of the (6,6) armchair
SWCNT\cite{Krompiewski04jmmm} due to the almost perfect fitting of
the armchair nanotube lattice constant ($a=2.49$~\AA) to the
interatomic distances at the electrode close packed surfaces
(\textit{e.g.}~2.51, 2.49 and 2.55~\AA\ for Co, Ni and Cu
respectively).  This coincidence makes it possible to put the
interface (contact) ring of carbon atoms on top of the electrode
substrate in such a manner that all carbon atoms sit in the
geometrical centers of three adjacent electrode-atoms, and
additionally the perimeter of the contact carbon ring equals,
within a few percent, the standard value for the (6,6) SWCNT,
\textit{i.e.}~$6\sqrt{3}a$. On geometrical grounds, the only other
armchair structure which can be constructed likewise is the (2,2)
one (see Fig.~\ref{fig:fig1}).  So we will use it in our studies
as the inner tube,
 in spite of the fact that
it is known to be energetically
unstable.\cite{CharlierM93+QinZHMAI00} As a matter of fact, since
we focus here on electrical transport properties rather than
stability problems,  we are convinced that the present model gives
a qualitative insight into the role of the inter-wall interaction
in real MWCNTs, depending on whether or not the inner tubes are
fully contacted to the electrodes.  In this very context, the
present model is advantageous in view of its small number of atoms
within the unit cell, ideal geometrical fitting to the electrode
surface atoms, and its well-defined inter-wall coordination zones.
The last feature enables us to define inter-wall hopping integrals
in such a way that each inner-tube carbon atom has just two
outer-tube atoms to interact with.  Thus, again no ambiguity
exists concerning the interaction range, and no hidden cutoff
parameters are needed. Fig.~1 corresponds to an odd number of
carbon rings (inverse-symmetric system). The case of even numbers
of rings makes no serious complication\cite{Krompiewski04jmmm},
but the drain electrode has to be rotated by $\pi/3$ with respect
to the source one if the perfect geometrical matching at the drain
interface is also required.

\begin{figure}[t]
\centerline{\includegraphics[width=.99\linewidth]{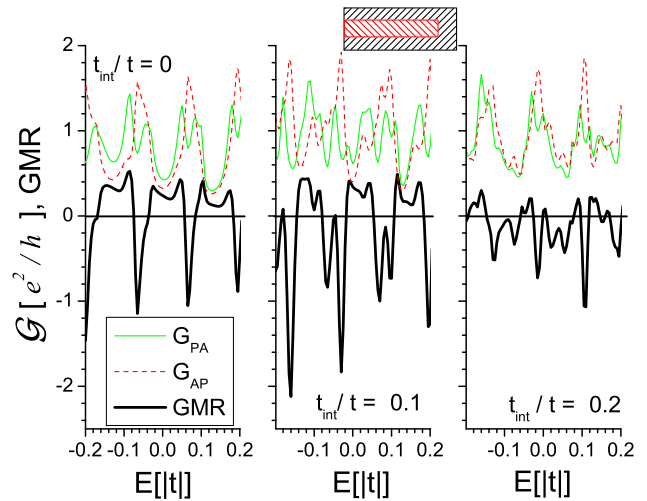}}
\caption{\label{fig:fig3} (Color online) Unlike in the previous figures, here the
inner tube is out of contact to the right electrode (see the upper
right diagram).  A drastic drop in the conductance (at $E_{\rm F}$),
accompanied by a positive GMR, takes place.  No inter-wall
interactions (left panel), and $t_{\rm int}/t = 0.1$ and $0.2$
(middle and right panels, respectively). }
\end{figure}
\section{Formalism}
 The device under consideration
 is described by means of
a single-band tight-binding Hamiltonian, assuming
$\pi$ and $s$-electrons
in the double-wall carbon
nanotube and ferromagnetic-electrodes, respectively. The total Hamiltonian can
be compactly written as
\begin{equation} \label{hamiltonian}
H = \sum \limits_{\lab i ,j \rab}
t_{i,j}\labs i \rab \lab j \rabs
+\sum \limits_{i}
\epsilon_i \labs i \rab \lab i \rabs,
\end{equation}
with vanishing initial on-site potentials, $\epsilon_i$, in the
DWCNT, whereas in the electrodes, they are spin-dependent and
chosen so as to give the required magnetization.  To solve the
Green function problem, we use the partitioning
technique,\cite{DamleRPD02} treating the whole device as a
left-electrode -- DWCNT -- right-electrode system (shorthand
notation L-C-R).  Here C stands for the central carbon-based part,
which incorporates not only the DWCNT but also the first two
closest atomic layers (with $N=36+49$ atoms) from both the L- and
R-electrode, implementing the extended molecule concept (as shown
in Fig.~\ref{fig:fig1}).  The Green function of the extended
molecule reads
\begin{equation} \label{eq:G}
G =\lrb  E - H_{\rm C}-\Sigma_{\rm L}-\Sigma_{\rm R}\rrb ^{-1},
\end{equation}
whereas the density matrix and the conductance (per spin) are
given by:
\begin{figure}[t]
\centerline{\includegraphics[width=.99\linewidth]{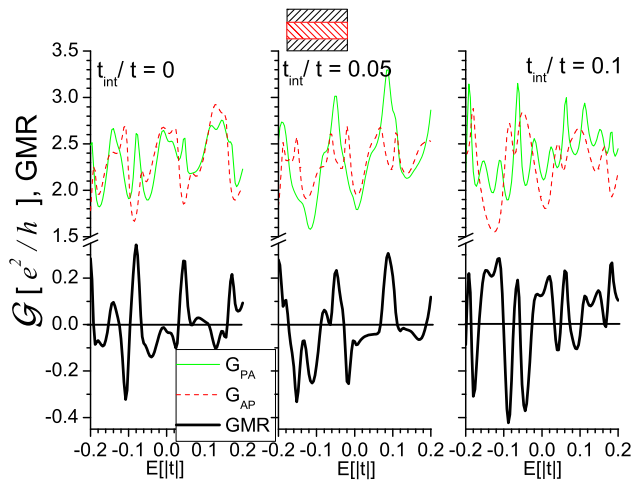}}
\caption{\label{fig:fig4} (Color online) As Fig.2 but for 40 carbon rings in each
nanotube. }
\end{figure}
\begin{equation} \label{eq:rho}
n=
\frac{1}{2 \pi} \int dE \, G \lrb f_{\rm L} \Gamma_{\rm L}+f_{\rm R} \Gamma_{\rm R} \rrb G^\dagger,
\end{equation}
\begin{equation} \label{I}
{\cal G}= \frac{e^2}{h} \textrm{Tr}\, \lcb \Gamma_{\rm L} \;G\;\Gamma_{\rm R} \;G^\dagger \rcb,
\end{equation}
with
$\Gamma_\alpha = \ii ( \Sigma_\alpha -
\Sigma_\alpha^\dagger ), \, \, \, \, \Sigma_\alpha=V_{{\rm
C},\alpha} \, g_\alpha \, V_{{\rm C},\alpha}^\dagger \, \,$ and
$f_\alpha$ being the Fermi-Dirac distribution function
($\alpha=$~L, R).
The $\alpha$-th electrode surface Green function, $g_\alpha$, is
calculated as described in Ref.~\onlinecite{TodorovBS93}. While
back Fourier transforming to the real space, the integration over
the two-dimensional Brillouin zone has been performed by the {\it
special-k-points method}.\cite{Cunningham74} The giant
magnetoresistance is defined as GMR$= ({\cal
G}_{\uparrow,\uparrow}-{\cal G}_{\uparrow,\downarrow})/{\cal
G}_{\uparrow,\uparrow}$, where the arrows denote the aligned and
antialigned magnetic configuration.  To parameterize the
Hamiltonian, we have put the polarization of the electrodes equal
to 50\% (with the number of electrons per atom being 0.75 and 0.25
for majority and minority bands, respectively).  The energy scale
is given by the hopping integral $|t|=2.66$~eV and the nanotube
lattice constant is $a=2.49$~\AA , fixing the energy and length
units, respectively. In order to minimize the number of free
parameters the nearest neighbor hopping integral is kept constant
in the entire device. As regards the inter-tube interaction we
treat the hopping integral as a free parameter of the order of
$t_{int} \sim t/10$ (as  for the interlayer distance in
graphite\cite{Ferreira01}), although in general $t_{int}$ depends
exponentially on distance and also on angles between the $\pi$
orbitals of the involved atoms.\cite{Lambin00}
It should be also admitted that another complexity, not captured
by our simple structural model (two armchair tubes), is a possible
additional off-diagonal disorder due to incommensurability between
shells (cf. Ref. \onlinecite{Roche01}).\\

\begin{figure}[t]
\centerline{\includegraphics[width=.99\linewidth]{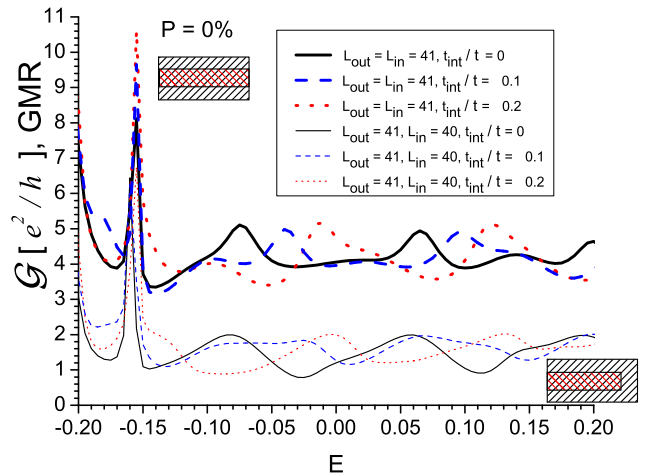}}
\caption{\label{fig:fig5} (Color online) The conductance in the case of
paramagnetic-electrodes. The breakage of the inner-tube contact to
one of the electrodes results in a drastic
 drop of the conductance (the lower curves).  The inter-wall hopping integrals,
  $t_{\rm int}/t$, are equal to $0, 0.1$ and $0.2$ (solid, dashed and dotted curves, respectively).
}
\end{figure}
\section{Results}
Our energy band line-up procedure goes as follows: First, $E_{\rm F}$ is
fixed at 0, next the on-site potentials in the electrodes are set
so as to yield the required number of electrons per atom ($n=1$) and
magnetic moment (0.5 and 0 in the ferromagnetic and paramagnetic
cases, respectively). The so-determined Fermi energy is kept fixed at the charge
neutrality point and no doping effects are taken into account. 
Finally, all the on-site potentials within
the extended molecule are self-consistently computed from the
global charge neutrality condition and from Eq.~(\ref{eq:rho}). Rather
than restrict ourselves  to the Fermi energy, we present our
results in an energy window, accessible e.g. by applying a gate
potential (see e.g Ref.~\onlinecite{KrompiewskiMB02}).

We have first considered a perfectly contacted case, when both
walls are in touch with the source and drain electrodes.  The
results, presented in Fig.~\ref{fig:fig2}, show that for the MWCNT
length equal to 20.5{\it a} (41 carbon rings) the GMR is negative
regardless of whether inter-wall interactions are included or not.
Next, the inner tube has been allowed to be out of contact to,
say, the drain electrode, see Fig.~\ref{fig:fig3}.  It turns out that then, not only the
conductance of the whole system decreases roughly twice, but also
the GMR becomes positive.  The effect of the inter-wall
interactions is moderate (at least up to $t_{int}/t=0.1$), they
happen to increase the conductance, always keeping the GMR positive.
The inverse GMR effect in the case of full end-contacted
nanotubes, and $t_{\rm int}/t$ not exceeding a length-dependent
threshold value, has been also found for the nanotube lengths of
40- and 39-rings, and is depicted in Fig.~\ref{fig:fig4} for $N$=40.
As it is seen, the GMR is now far less robust than for $N$=41.
Judging from the conductance behavior at $E_{\rm F}$ for 3 consecutive
nanotube lengths, one can anticipate what happens for an arbitrary
length, because the conductance is a quasi-periodic function with
a period equal to 3-rings spacing. Thus, for DWCNTs which are $3N-1$
rings long we predict a quite pronounced negative GMR effect ($N$ is
an integer). The $3N-1$ length rule (in a/2 units) was numerically
found\cite{Krompiewski03pss,Krompiewski04jmmm}, with these lengths
an ideal nanotube has got maximum conductance (''on-resonance
device''). Disorder at the interfaces makes the maximum of the
conductance shift beyond $E_{\rm F}$, but the quasi-periodicity is still
preserved. The aforementioned quasi-periodicity originates from the fact that
ideally the SWCNT armchair energy spectrum crosses the
[$k=2\pi/(3a), E=0$]-point for particular lengths only, and
additionally conductance is a measure of the squared electron wave
function.

As a test illustrating the quality of the present approach, 
we have completed our studies by additional computations for
DWCNTs sandwiched between \textit{paramagnetic} electrodes with the
same number of electrons per lattice site ($n = 1$) as in the
ferromagnetic calculations. Our attention has been again focused
both on the effect of the inner nanotube contact to one of the
electrodes, and the importance of the inter-wall interactions. The
latter are now allowed to take values up to $t_{\rm int}/t=0.2$.
Fig.~\ref{fig:fig5} shows that the conductance in the vicinity of
$E_{\rm F}$ does not reach the maximum theoretical value of $8
e^2/h$ for a DWCNT ($4 e^2/h$ contribution for every wall obtained
for infinite MWCNTs). The conductance suppression in MWCNTs below
the expected value for the infinite homogeneous tube has been
experimentally well documented.\cite{FrankPWdH98,PoncharalFWdH99}
In our model system charge-transfer induced changes in the band
structure of the DWCNT lead to a reduction (roughly to $4e^2/h$)
of the conductance -as one can see in the upper curves in
Fig.~\ref{fig:fig5}. This finding is in line with the well-known
scenario, that  the reduction of the conductance is due to the
interface-induced suppression of one of the two transport channels.\cite{KwonT98+SKTL00} 
Obviously, if a finite armchair
nanotube is contacted to external electrodes it can no longer be
viewed as a periodic one with the repeat unit consisting of two
carbon rings, so the band degeneracy is lifted and the two channels
couple very differently to their counterparts in
the leads. It is noteworthy that  recent {\it ab initio} results on
single wall carbon nanotubes end-contacted to paramagnetic
electrodes also show that the interface mismatch may result in
roughly halving the conductance.\cite{Palacios03}

The lower part of Fig.~\ref{fig:fig5} shows the effect of the
inner-tube contact breakage.  The conductance gets reduced by a
factor close to two, very much like in the ferromagnetic
computations.  More importantly however, this behavior is in
qualitative agreement with experimental
results\cite{FrankPWdH98,PoncharalFWdH99} if we associate the
lower and upper bunch of curves in Fig.~\ref{fig:fig5} with the
major plateaux of the conductance of MWCNT immersed into Hg.
Incidentally, the peculiar peak in Fig.~\ref{fig:fig5} more than
0.15 $\labs t \rabs $ (ca. 0.4 eV) away from the charge neutrality
point, is due to model-specific features of no universal nature.

Both the GMR and the conductance are oscillatory, non-monotonic
functions of the inter-wall coupling, and reveal a quasi-periodic
behaviour with the DWCNT length. It is also noteworthy that the
conductances in the ferromagnetic case are always clearly smaller
(regardless of the electrode magnetization alignment) than their
paramagnetic counterparts.  We attribute this behavior to the
increased energy band mismatch at the interfaces in the
ferromagnetic case.

\section{Conclusions}
In conclusion, we have investigated the effect of inter-wall
interactions on the giant magnetoresistance of double wall carbon nanotubes,
 depending on whether or not the inner
wall is fully contacted to the electrodes.
We have found
that the GMR of perfectly contacted DWCNTs is  very
sensitive to the inter-wall coupling strength, and show
a tendency to become negative.  On the contrary, DWCNTs
with inner tubes being out of contact with one of the
electrodes reveal (close to the Fermi energy) a robust positive GMR only weakly dependent on the
inter-wall coupling.  These observations may be related
to the experimental results showing either positive or
negative (sample-dependent) GMR of comparable value ($\sim 30\%$),\cite{ZhaoMVMS02}
 as well as recently reported negative GMR values in
MWCNTs strongly contacted to permalloy
electrodes.\cite{KimKPKKKW02} Moreover, our complementary
calculations for the case with paramagnetic electrodes show that
the conductance of the perfectly contacted DWCNT is suppressed by
a factor two with respect to the case of a homogeneous infinite
DWCNT (that is $4e^2/h$ instead of $8e^2/h$). Consequently, for
the case in which the inner wall gets disconnected, we observe
conductance values close to $2e^2/h$ instead of $4e^2/h$. This
is related to the
charge transfer induced energy band rearrangements in the extended
molecule. The present approach may be regarded as a reference for
further generalizations. The diameters and the number of walls
forming the MWCNT may be easily increased and the respective
stable positions of the interface atoms can be readily determined
by a simple relaxation procedure involving just a few variational
parameters (displacements and rotations).

\begin{acknowledgments}
We are indebted to Gerrit E. W. Bauer, Uwe Krey and Nitesh Ranjan for fruitful
discussions.  SK acknowledges the support from the DAAD Foundation
and the KBN research project PBZ-KBN-044/P03-2001.  This work was
partially funded by the Volkswagen Stiftung and the EU Centre of
Excellence contract G5MA-CT-2002-04049.
\end{acknowledgments}

\end{document}